\documentclass[revtex4]{emulateapj}
\usepackage{graphicx, pdflscape, amsmath, natbib,color}
\usepackage[FIGTOPCAP]{subfigure}

\usepackage{savesym}
\savesymbol{tablenum}
\usepackage{siunitx}
\restoresymbol{SIX}{tablenum}
\usepackage{longtable}
\usepackage{threeparttablex}
\usepackage{booktabs}

\begin{document}

\defcitealias{2015PNAS..112.8965B}{B15}

\title{Destruction of Refractory Carbon in Protoplanetary Disks}
\author{Dana E. Anderson\altaffilmark{1}, Edwin A. Bergin\altaffilmark{2}, Geoffrey A. Blake\altaffilmark{1}, Fred J. Ciesla \altaffilmark{3}, Ruud Visser \altaffilmark{4}, Jeong-Eun Lee \altaffilmark{5} }

\affil{$^1$Division of Geological and Planetary Sciences, California Institute of Technology, 1200 E. California Blvd., Pasadena, CA 91125, USA}
\affil{$^2$Department of Astronomy, University of Michigan, 1085 S. University, Ann Arbor, MI 48109-1107, USA}
\affil{$^3$Department of Geophysical Sciences, The University of Chicago, 5734 South Ellis Ave., Chicago, IL 60637, USA}
\affil{$^4$European Southern Observatory, Karl-Schwarzschild-Str. 2, D-85748, Garching, Germany}
\affil{$^5$School of Space Research, Kyung Hee University, 1732, Deogyeong-daero, Giheung-gu, Yongin-si, Gyeonggi-do 17104, Korea}

\begin{abstract}
\noindent
The Earth and other rocky bodies in the inner solar system contain significantly less carbon than the primordial materials that seeded their formation. These carbon-poor objects include the parent bodies of primitive meteorites, suggesting that at least one process responsible for solid-phase carbon depletion was active prior to the early stages of planet formation. Potential mechanisms include the erosion of carbonaceous materials by photons or atomic oxygen in the surface layers of the protoplanetary disk. Under photochemically generated favorable conditions, these reactions can deplete the near-surface abundance of carbon grains and polycyclic aromatic hydrocarbons by several orders of magnitude on short timescales relative to the lifetime of the disk out to radii of $\sim$20--100+~au from the central star depending on the form of refractory carbon present. Due to the reliance of destruction mechanisms on a high influx of photons, the extent of refractory carbon depletion is quite sensitive to the disk's internal radiation field. Dust transport within the disk is required to affect the composition of the midplane. In our current model of a passive, constant-$\alpha$ disk, where $\alpha$ = 0.01, carbon grains can be turbulently lofted into the destructive surface layers and depleted out to radii of $\sim$3--10~au for 0.1-\SI{1}{\micro\metre} grains. Smaller grains can be cleared out of the planet-forming region completely. Destruction may be more effective in an actively accreting disk or when considering individual grain trajectories in non-idealized disks.

\end{abstract}

\section{Introduction}
The earliest stages of planet formation begin within a primordial disk of gas and submicron-sized dust grains surrounding a young star, formed due to the conservation of angular momentum through the collapse of a dense core in a molecular cloud. This material inherited from the interstellar medium (ISM) experiences a range of physical conditions and chemical environments. The dust grains within the disk concentrate toward the midplane due to the gravitational pull from the central star over time, increasing their likelihood of interaction. Dust grain aggregates are thought to grow through collisions, forming bodies that are orders of magnitude larger in size than typical interstellar dust as part of the planet formation process \citep[e.g.][]{1993prpl.conf.1031W}. 

Provided that the original grains are of approximately interstellar composition, the amount of carbon available at the onset of planet formation can be roughly estimated. The carbon-to-silicon abundance ratio of the ISM is similar to that of the solar photosphere, C/Si$\sim$10 (\citealp{2015PNAS..112.8965B}, hereinafter B15; \citealp{2010Ap&SS.328..179G}).  About 60\% of carbon in the ISM is present in an unidentified solid, and thus potentially refractory, form \citep{1996ARA&A..34..279S}. Suggested components include graphite, amorphous carbon or hydrocarbon grains with a combination of aliphatic-rich and aromatic-rich components \citep{2013ApJ...770...78C, 2013A&A...558A..62J}, large organics, and/or polycyclic aromatic hydrocarbons (PAHs). Once in the protoplanetary disk, carbon retained in the gas phase is subject to accretion by the central star or gas-giant planets or else to dissipation, ultimately being largely cleared from the disk. Any remaining volatile carbon species would likely be found in the form of ices in only the cold outer regions of the disk. In contrast, one would expect more-refractory carbon sources with vaporization temperatures of 425--626~K \citep{1994ApJ...421..615P, 2003ApJ...591.1220L} to be readily available and incorporated into solid planetary bodies, including planetesimals, throughout the entire disk except for the very innermost radii \citep[$\lesssim$0.5~au from the central star,][]{2005ASPC..341..353D} and the directly irradiated surface. However, observations of rocky bodies in our solar system reveal a significant depletion in carbon compared to the progenitorial interstellar dust (see Table~\ref{ctosi}; \citealp{1987A&A...187..859G, 2010ApJ...710L..21L, 2014prpl.conf..363P}; \citetalias{2015PNAS..112.8965B}).

The C/Si ratio of the bulk silicate Earth (BSE; {the entire Earth, including the oceans and atmosphere, excluding the core}) is about four orders of magnitude lower than that of the interstellar dust that seeded its formation \citepalias{2015PNAS..112.8965B}. This depletion of carbon could be the result of various events throughout Earth's formation and early stages, including devolatilization during processing of accreted material or impact events and sequestration of carbon via core formation (see \citetalias{2015PNAS..112.8965B} and references therein). However, carbonaceous chondrites, some of the most primitive materials known in our solar system, also exhibit carbon depletions \citep[by one to two orders of magnitude;][]{1988RSPTA.325..535W} relative to the interstellar value. This suggests that a significant amount of the refractory carbon, which would have otherwise been incorporated into these solid bodies at a level of $\sim$6$\times$ that of silicon, was destroyed early in the solar system's history and prior to the formation of rocky planets and differentiated planetesimals. 

Not all solid bodies in our solar system are carbon poor. The dust of comet Halley has a C/Si ratio similar to that of the ISM (\citealp{2014prpl.conf..363P}; \citetalias{2015PNAS..112.8965B}) and anhydrous interplanetary dust particles (IDPs), argued to be of cometary origin \citep{2006mess.book..187M}, also have high carbon content. The increased carbon retention by these outer-solar-system bodies cannot be explained by volatile ices alone; at least 50\% of the carbon in comets is thought to be in refractory form \citepalias{2015PNAS..112.8965B}. Unlike Halley, the Sun-grazing comets resemble carbonaceous chondrites in terms of carbon abundance. These differences suggest that the mechanism responsible for refractory carbon destruction did not operate uniformly throughout the solar system and may have been less active in some comet-forming regions.   

Evidence for carbon-deficient bodies is also found beyond our solar system. Spectra of polluted white dwarf atmospheres provide an opportunity to view the chemical composition of extrasolar planestimals. The strong gravitational forces of white dwarfs are expected to deplete heavier elements from their photospheres on relatively short timescales. Therefore, heavy elements observed in the atmospheres of white dwarfs are believed to have originated in rocky bodies that are destroyed near the tidal radius and accreted by the star \citep{2007ApJ...671..872Z, 2015MNRAS.447.1049V}. Elemental measurements reveal these bodies to be carbon-poor with C/Fe ratios similar to those measured in chondritic meteorites in our own solar system \citep{2014AREPS..42...45J, 2016MNRAS.463.3186F, 2016MNRAS.459.3282W}.

Explaining the C/Si record of solid bodies in the solar system requires a mechanism that (1) selectively removes refractory carbon from the condensed phase while leaving silicates intact, (2) operates prior to the formation of the parent bodies of meteorites, (3) accounts for a range in carbon content among different solar-system bodies, and (4) may be a common process that is not unique to the specific conditions of our solar system. Once the disk enters the passive state, meaning that it is no longer accreting material from the interstellar cloud and is heated mainly by irradiation from the central star, vaporization of refractory carbon will be ineffective throughout most of the disk because the dust temperatures are too low \citep{2005ASPC..341..353D}. 

Alternative destruction mechanisms include those driven by energetic radiation. For example, photochemically activated oxidation of carbonaceous material via chemical reactions with O-bearing species, including OH and free atomic O, which can erode the surface of carbon grains releasing carbon into the gas phase \citep{1997A&A...317..273B, 1997A&A...325.1264F, 2001A&A...378..192G, 2002A&A...390..253G}. \cite{2010ApJ...710L..21L} calculated that combustion through reactions with atomic O, which is abundant in the photochemically active surface layers of the disk, can efficiently destroy carbon grains $<$\SI{10}{\micro\metre} in size and given sufficient vertical transport in the disk, potentially explain the observed carbon deficiencies of rocky bodies in our solar system. Direct  photochemical destruction of refractory carbon sources, resulting in the ejection of small hydrocarbons, represents an additional destruction mechanism. \citet{2014A&A...569A.119A, 2015A&A...584A.123A} experimentally
\begin{ThreePartTable}
\begin{TableNotes}
\footnotesize
\item \textsc{Note.}---See \citet{2015PNAS..112.8965B} and sources therein
\end{TableNotes}
\begin{longtable}[!b]{lc}
\caption{\\ \textsc{Bulk C/Si in Solar-system Bodies}}\\
\hline\hline
\vspace{-1.8 mm} \\ 
Source & C/Si \\
\midrule
\endhead
\midrule
\insertTableNotes
\endfoot
ISM (dust) & 6 \\
Earth (BSE) & 0.001 \\
Meteorites (CI, CM) & 0.4--0.7 \\
Meteorites (CO, CV) & 0.07--0.08 \\
Comets (Halley, dust) & 6 \\
Comets (Halley, gas + dust) & 8 \\
Comets (Sun-grazing) & 0.08--0.2 \\
Interplanetary Dust Particles (IDPs) & 2 
\label{ctosi}
\end{longtable}
\end{ThreePartTable}
\noindent  investigated the rate of vacuum ultraviolet (VUV) photolysis for interstellar hydrogenated amorphous carbon (HAC) analogs; whereas the ability of large aromatic carbon species to survive photodissociation has been evaluated through modeling by \citet{2007A&A...466..229V}. \citet{2010A&A...511A...6S} and \citet{2012A&A...543A..25S} also explored the potential for PAH destruction via more energetic extreme ultraviolet (EUV) and X-ray photons.

Here we expand upon the work of \cite{2010ApJ...710L..21L}, exploring the possibility that refractory carbon sources in the disk can be depleted before grains grow to sufficient sizes and while most of the solid mass remains in 0.1--\SI{1}{\micro\metre} baseline seed particles. We use a full chemical disk model to explore the oxidation and photochemical destruction of two sources of refractory carbon: carbon grains and PAHs. With this model, we attempt to estimate the efficacy of these mechanisms throughout the disk and determine the radial distance to which carbon can be depleted for comparison with our own solar system. Section 2 describes the disk model used in our analysis, while Section 3 provides the results of the model and subsequent analysis of the capabilities of the proposed destruction mechanisms. A discussion of the results is presented in Section 4 followed by a final summary in Section 5. 

\begin{figure*}[t]
\centering
\includegraphics[scale=1.1]{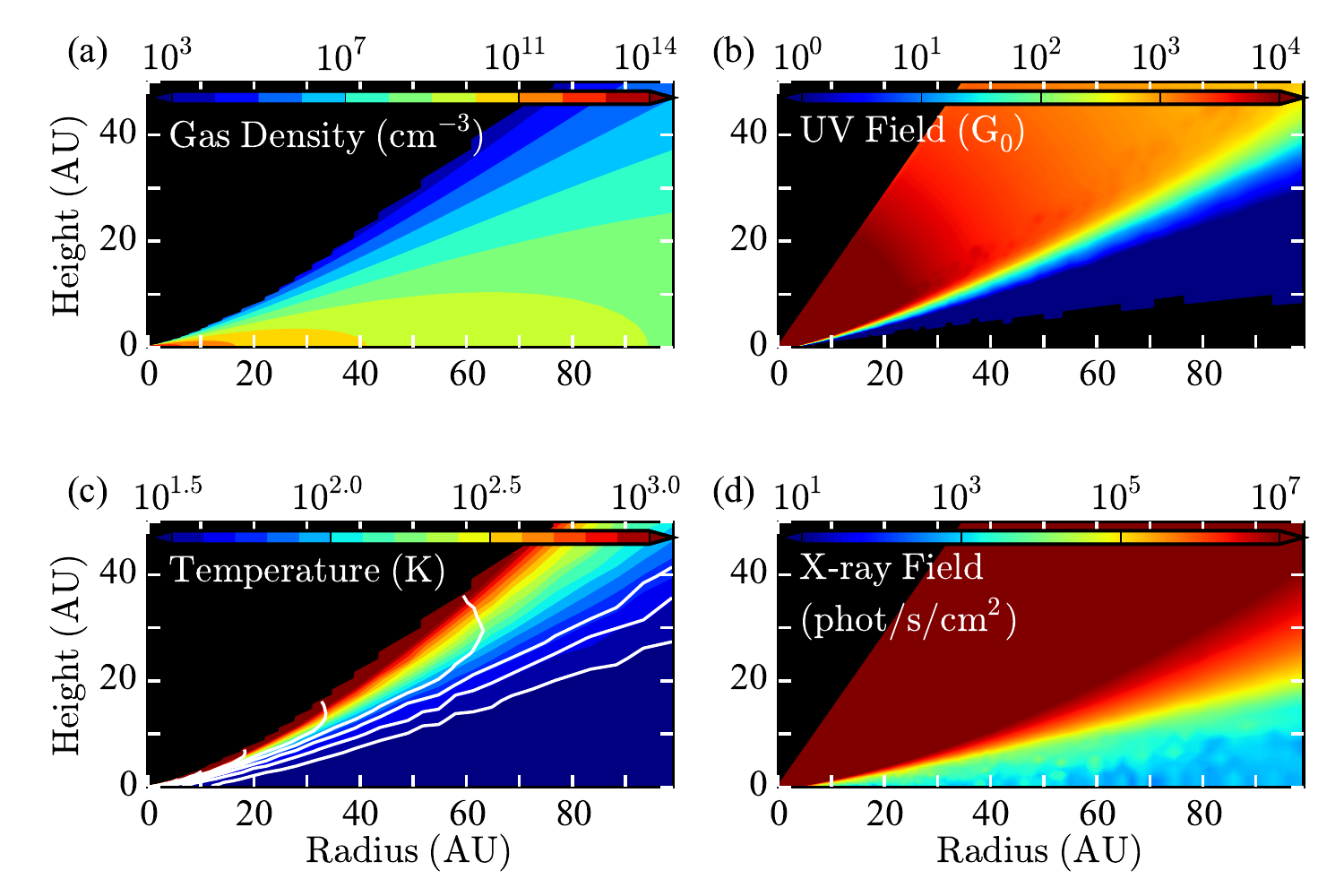}
\caption{Physical conditions for our disk model: gas density (a), integrated UV flux, 930--\SI{2000}{\angstrom} (b), gas temperature with dust temperature shown in white contours increasing in intervals of 10$^{0.1}$\SI{}{\kelvin} from 10$^{1.5}$\SI{}{\kelvin} in the midplane (c), and integrated X-ray flux, 1--\SI{20}{\keV} (d) out to 100~au from the central star.}
\label{Disk}
\end{figure*}

\section{Model}
Here we model a static, passive protoplanetary disk around a T-Tauri star, a young, low-mass star thought to be similar to our early Sun. Chemistry within the disk is predicted by solving a set of ordinary differential equations describing the chemical reaction rates of the system. To the standard chemical model, we add a source of refractory carbon, either carbon grains or PAHs, and appropriate destruction mechanisms in order to track the abundance of these sources throughout the disk.

\subsection{Physical Conditions}
To provide the conditions representative of a disk surrounding a typical T-Tauri star, we use a physical disk model similar to that of \cite{2013ApJ...772....5C, 2015ApJ...799..204C}. The model is two-dimensional and azimuthally symmetric. To approximate the dust distribution of an early disk, prior to substantial grain growth and sequestration of mass in the midplane, we include a single population of small dust (radii $\leq$1--\SI{10}{\micro\metre}) that follows the distribution of the gas. The dust composition is 80\% astronomical silicates \citep{1984ApJ...285...89D} and 20\% graphite. Although ice coatings on the dust may affect the resulting opacity \citep{1994A&A...291..943O}, the model assumes the general approximation of bare grain surfaces. Given our static model, these grains are considered only for calculating the physical conditions of the disk and are not directly included in the chemical reactions. This dust distribution therefore remains unchanged as we destroy carbon-bearing grains. The gas and dust share a fixed radial surface density profile described by a power law with an exponential cutoff, $\Sigma_{gas}(R) = \Sigma_{c} (R/R_{c})^{-1}exp[-(R/R_{c}$)] where $\Sigma_{c}$ =  3.1~g~cm$^{-2}$ and $R_{c}$ = 135~au from \cite{2013ApJ...772....5C}, and are related by assuming a vertically integrated gas-to-dust mass ratio of 100.  The total mass of the disk gas is 0.039 M$_{\odot}$.

The UV field, including the propagation of Ly $\alpha$ photons, and the X-ray radiation field within the disk were generated using a Monte Carlo radiative transfer code as described by \citet{2011ApJ...740....7B,2011ApJ...739...78B}. The input UV spectrum is taken to be that observed for TW Hya \citep{2002ApJ...572..310H, 2004ApJ...607..369H}.  The modeled X-ray spectrum has a total luminosity of 10$^{29.5}$ erg s$^{-1}$ between 1--20 keV, corresponding to that of a typical T-Tauri star. The gas temperature is estimated based on the computed UV field and gas density distributions using the thermochemical models of \cite{2013A&A...559A..46B}, as described in \cite{2015ApJ...799..204C}. Figure~\ref{Disk} shows the final density, temperature, and radiation distributions provided to the chemical model.
\vspace{1cm}
\subsection{Chemical Network}
The disk chemistry model combines the higher-temperature ($\lesssim$\SI{800}{\kelvin}) gas-phase network of \cite{2010ApJ...721.1570H} with the gas-grain network of \cite{2011ApJ...726...29F}. In total, over 600 chemical species and nearly 7000 reactions are specified, including gas-phase reactions among ions, neutrals, and electrons; photodissociation; X-ray and cosmic-ray ionization; gas-grain interactions (freeze-out, sublimation, photodesorption, and cosmic-ray-induced desorption); and formation of OH, H$_2$O, and H$_2$ on grains. The model is run for radii from 0.25 to 100~au from the central star, divided into a grid of 42 logarithmically spaced radii each containing 45 linearly spaced vertical zones. With the exception of H$_2$ and CO self-shielding, which are considered throughout the vertical column, chemistry within each grid space is modeled independently. The disk model is static, with no mixing between grid spaces. To this chemical network, we added sources of refractory carbon as described in the following sections.

\subsection{Sources of Refractory Carbon}
\subsubsection{Carbon Grains}
The first source of refractory carbon included in our model is solid grains. Different grain sizes are considered, with radii of 0.01--\SI{10}{\micro\metre}. Sizes are larger than the average interstellar values to account for growth within the disk. Large, $\sim$1--\SI{10}{\micro\metre}, silicate grains have been observed in emission at 10 and \SI{18}{\micro\metre} originating from the irradiated layers. Modeling the grains as having compact, spherical structures provides an upper limit on the destruction timescale. However, during growth due to collisions the aggregate structures can become open. Laboratory experiments using dust analogs find that initial growth through collisions resulting in the sticking and freezing together of individual particles leads to the formation of fractal structures \citep{2008ARA&A..46...21B}. In addition, interplanetary dust particles collected from Earth's atmosphere are described as typically being porous aggregates composed of $\sim$10$^6$ grains \citep{1979RvGSP..17.1735B}. To account for porosity in our modeled carbon grains, volume-enlargement factors from the collisional evolution simulations of \cite{2007A&A...461..215O} are used. The resulting porosities are assumed to be 0\% for the 0.01 and \SI{0.1}{\micro\metre} grains, 88\% for the \SI{1}{\micro\metre} grains, and 96\% for the \SI{10}{\micro\metre} grains. The porosity ($\phi$) represents the percentage of the grain volume that is unoccupied. Compared to a compact configuration of atoms, a porous grain of the same size will only contain (100-$\phi$)\% of the amount of carbon.

Oxidation of these carbon grains is modeled based on the approach of \cite{2010ApJ...710L..21L}. An oxygen atom that collides with a carbon grain removes a single C atom to form gaseous CO. Oxidation by OH is also possible \citep{1997A&A...317..273B, 1997A&A...325.1264F, 2001A&A...378..192G, 2002A&A...390..253G}, but O is more abundant in the photoactive surface layers of the disk. 
The reaction occurs at a rate of $R_{ox} = n_{cgr} n_{o} \sigma v_{o} Y$, where $ n_{cgr}$ is the number density of carbon grains, $n_{o}$ is the number density of oxygen atoms, $\sigma$ is the cross-section of a carbon grain, $v_{o}$ is the thermal velocity of an oxygen atom, and $Y$ is the yield of the reaction (\citealp{1979ApJ...230..106D}, $Y = 170\exp(-4430/T_{gas})$ if $T_{gas} > $ \SI{440}{\kelvin} and $Y = 2.30\exp(-2580/T_{gas})$ if $T_{gas} \le $ \SI{440}{\kelvin}). The abundance of carbon grains is determined by taking the initial abundance of refractory carbon and dividing it by the number of carbon atoms per grain, equal to $[\rho_{cgr} \frac{4}{3} \pi r^3 (100\%-\phi)]/m_{c}$ where $\rho_{cgr}$ is taken to be the density of graphite, 2.24 g cm$^{-3}$; $r$ is the grain radius; $\phi$ is the porosity; and $m_{c}$ is the mass of a carbon atom. 

Photolysis rates for carbon grains of amorphous structure are taken from \citet{2014A&A...569A.119A, 2015A&A...584A.123A}, who measured the production rates of hydrogen and small hydrocarbons released from a plasma-produced HAC surface irradiated by VUV photons. Methane was the C-bearing product of highest yield and is the sole product considered here. The rate is $R_{UV}~=~n_{cgr} \sigma Y_{CH_4} (\Phi^{FUV}_{ISRF}/1.69) F_{UV}$, where $n_{cgr}$ and $\sigma$ are the same as above,  $Y_{CH_4}$ is the photo-production yield of CH$_4$ per incoming photon ($\sim$8$\times$10$^{-4}$), $\Phi^{FUV}_{ISRF}$ is the far ultraviolet (FUV) flux of the interstellar radiation field divided by 1.69 to convert from the Draine to Habing values, and $F_{UV}$ is the FUV flux relative to the interstellar value from the disk model. Here the term ``photolysis" is used to refer to the photon-induced release of small hydrocarbons from the surface of a grain containing $\sim$10$^{6}$ or more C atoms, distinct from the photodissociation of the PAHs, where the structure of a $\lesssim$100 C atom molecule is broken resulting in the loss of small hydrocarbons.

Small grains are also subject to destruction by higher energy, X-ray photons through heating to the point of sublimation or grain charging increasing electrostatic stress to the point of shattering. The conditions under which X-ray radiation will lead to grain destruction have been estimated by \cite{2001ApJ...563..597F} for emission from $\gamma$-ray bursts. X-ray emission from T-Tauri stars is dominated by 1--2 keV photons. The energy deposited by these photons will be insufficient to heat 0.01--\SI{10}{\micro\metre} grains above the sublimation temperature of refractory carbon throughout most of the disk \citep{2001ApJ...561..880N}. Shattering due to the buildup of electrostatic charge relies on successive interactions with X-ray photons. However, in the protoplanetary disk, grains will cycle between the surface and the midplane. Whereas X-ray ionization rates may dominate at the disk surface, in denser gas toward the midplane recombination with free electrons will be a competitive process potentially preventing the buildup of charge on grains. Cosmic rays and high-energy X-rays, if present in these dense layers, will be key sources of ionization in the inner disk resulting in the production of free electrons \citep{1999ApJ...518..848I, 2015ApJ...799..204C}. The susceptibility of silicate grains to shattering is slightly greater than that of carbonaceous grains \cite{2001ApJ...563..597F}. Therefore, this mechanism would not selectively destroy carbonaceous grains and efficient reformation of silicate grains would be required to explain the observed composition of solar-system bodies. Further modeling is required to understand the effectiveness of X-rays in reducing refractory carbon abundances.

\subsubsection{PAHs}
PAHs represent the second source of refractory carbon included in our model. Refractory carbon is introduced initially as large, 50 C atom PAHs representative of interstellar species. For comparison of destruction rates versus PAH size, additional models start with 20 C atom PAHs to demonstrate the breakdown of smaller PAHs after some initial destruction in the disk. As time progresses in the model, reactions with O, H, and OH dismantle the initial PAHs removing fragments containing $\sim$2 C atoms per reaction. Rates involving small PAHs from pyrene (16 C atoms) down to a single aromatic ring are from \cite{WF1997}. Where necessary, reverse-reaction rate coefficients are calculated using the thermodynamic data they provided.\footnote{We derive the reverse-reaction rate coefficient from $k_r=k_f/K$, where the forward reaction rate coefficient provided is $k_f$ = $A T^n\exp(-E/RT)$, and $A$, $n$, and the activation energy $E$ are given by \cite{WF1997}, $R$ is the gas constant, and $T$ is the gas temperature. The equilibrium constant is given by $K =\exp(- \Delta G/RT) =\exp(- \Delta H/RT + \Delta S/R)$ where $\Delta G$, $\Delta H$ and $\Delta S$ are the change in the standard Gibbs free energy, enthalpy, and entropy of the reaction.} For larger PAHs, rates are taken as those of the analogous reaction (O, H, or OH) with benzene as is the case for pyrene, phenanthrene, and naphthalene in \cite{WF1997}. Given that the specific structure of PAHs in astrophysical environments is largely unknown, the molecular structure of the large PAHs was not strictly considered aside from ensuring conservation of mass within the reaction network.

\citet{2007A&A...466..229V} provide UV photodissociation rates resulting in the loss of two C atoms for PAHs of different sizes based on theoretical calculations. These rates incorporate the modulated UV field, specifically the FUV field, at each grid location in the disk. C$_2$H is successively removed to break down the larger PAHs and reaction products from the photodissociation of PAHs smaller than pyrene were selected to mimic the oxidation reaction pathways where appropriate. 

The rate of photodissociation by X-ray photons follows \citet{2010A&A...511A...6S} using the cross-section for PAHs from \citet{2004A&A...414..123S}, N$_C$ $\times$ C$^{abs}$ where N$_C$ is the number of carbon atoms per PAH and C$^{abs} \simeq 10^{-17} \times (20/E)^{2.2}$ cm$^2$ is the cross-section per carbon atom for photon energies ($E$) greater than 20 eV. The energy-dependent cross section is multiplied by the X-ray flux distribution at $E$ integrated over $E$ = 1-20 keV. X-ray photons carry enough energy such that a single interaction can dismantle an entire 50 C PAH. In this case, we assumed the products of the reaction to be small hydrocarbons in our reaction network containing four or fewer C atoms. However, the photodestruction of large PAHs via X-rays has not been characterized through laboratory experiments and the dispersion of energy from X-ray photons among different molecular processes is not well known. Given the uncertainty in how to define an X-ray photodissociation rate for large PAHs \citep{1992MNRAS.258..841V, 2010A&A...510A..37M}, we also tested models where X-ray absorption resulted in the removal of C$_2$H (analogous to the network used for UV photodissociation of PAHs above) for reaction yields of 0.5 and 1.0.

The breakdown of PAHs with more than 60 C atoms may result in the production of fullerenes, such as C$_{60}$, that are highly stable and resistant to destruction. \citet{2015A&A...577A.133B} suggest that $\lesssim$1\% of the instellar PAH abudance may be in the form of C$_{60}$. PAHs in the surface layers will likely have predominately neutral charge. Abundant PAH cations are not produced by the radiation field of T-Tauri stars \citep{2007A&A...466..229V} and PAH anions will only be present in denser regions of the disk toward the midplane. The inclusion of PAH formation reactions is not important since we found the formation of benzene (as a proxy) to be inefficient in comparison to the rate of PAH destruction under the conditions in our disk model. 

\subsubsection{Model Setup}
At the onset of the model, carbon is divided between the refractory and volatile phases. Several iterations of the model were run with different initial forms for the refractory carbon. This phase is represented by either carbon grains of a single size and porosity (six different cases: $R$ = 0.01, 0.1, 1, or \SI{10}{\micro\metre},  and for the larger two sizes: porous or non-porous structures) or PAHs (two cases: initial PAHs of 50 or 20 C atoms in size). For the carbon grain scenarios, the total abundance of carbon \citep[$\sim$2 $\times$ 10$^{-4}$ relative to H, about the solar value;][]{2010Ap&SS.328..179G} is divided equally between the grains and the volatile phase. In the case of the PAHs, an abundance of 1.5 $\times$ 10$^{-5}$ C relative to H, approximately the abundance observed in the ISM \citep{2005pcim.book.....T}, is distributed among the 50 or 20 C molecules. Volatile species are set to 1 $\times$ 10$^{-4}$ relative to H. Nearly all the volatile carbon is initially stored in CO (with minor amounts in CN, C, HCN, C$^{+}$, HCO$^{+}$, H$_2$CO, and C$_2$H). All chemical species are spread uniformly in abundance relative to H over the entire grid of the disk at the beginning of the model run. Radial and vertical mixing within the disk are not considered here.

\section{Results} 

A chemical model of an irradiated (passive) protoplanetary disk is used to identify the regions where destruction of refractory carbon sources occurs. Further analysis of the timescales of grain transport and the distribution of grains within the disk allow for the estimation of the radial extent of refractory carbon depletion due to the mechanisms described here.

\subsection{Regions of Active Refractory Carbon Destruction}

\begin{figure*}[t]
\centering
\includegraphics[trim=0cm 1cm 0mm 0cm]{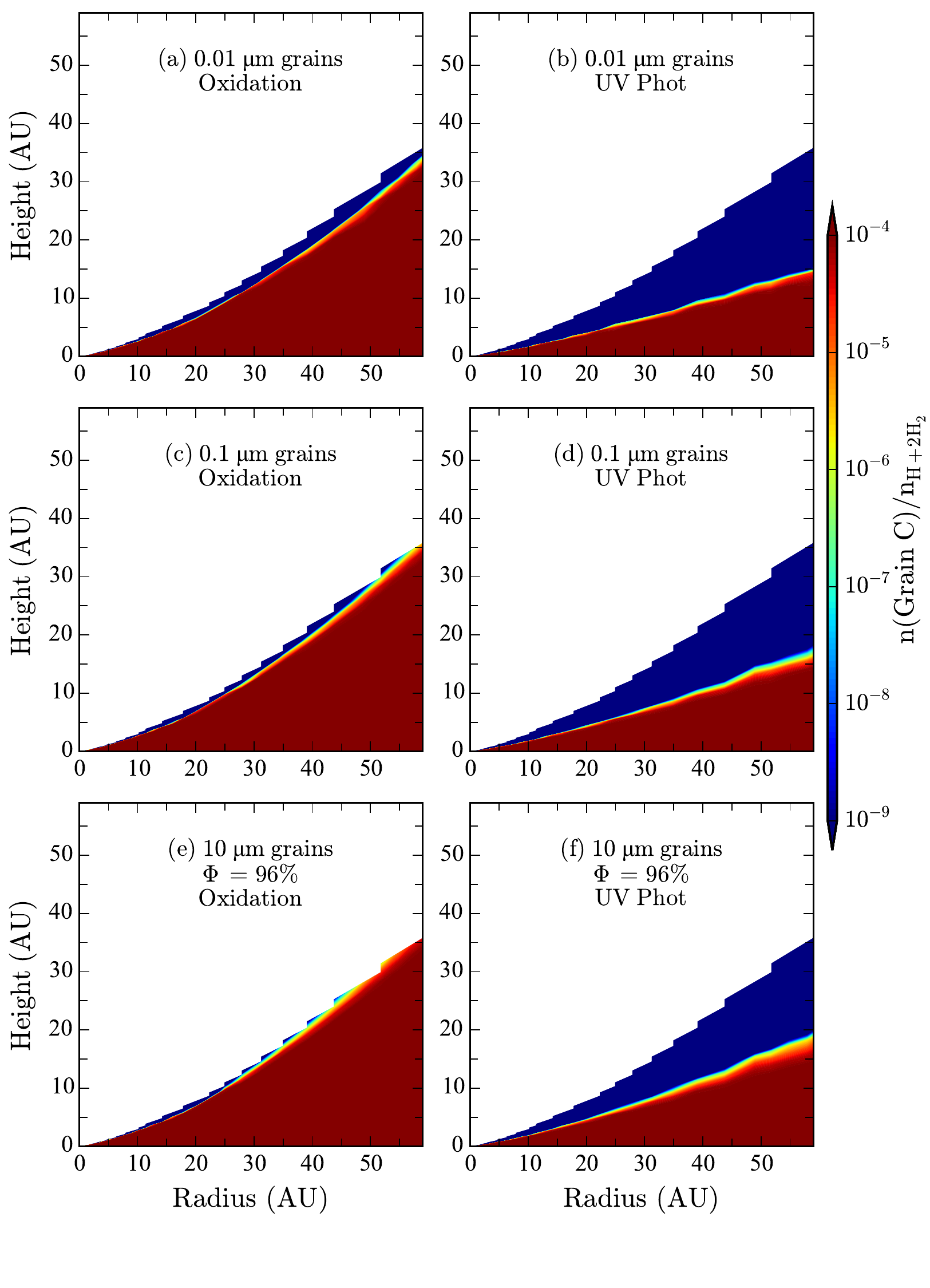}
\caption{Vertical cross-sections of the protoplanetary disk above the midplane demonstrating the refractory carbon abundance relative to total hydrogen at 10$^6$ years for carbon grains with grain radii equal to \SI{0.01}{\micro\metre} (a,b), \SI{0.1}{\micro\metre} (c,d), and \SI{10}{\micro\metre} with 96\% porosity (e,f) including oxidation (a,c,e) and UV photolysis (b,d,f).}
\label{FinalAbundance}
\end{figure*}

\begin{figure*}
\centering
\includegraphics[trim=0cm 1cm 0mm 0cm]{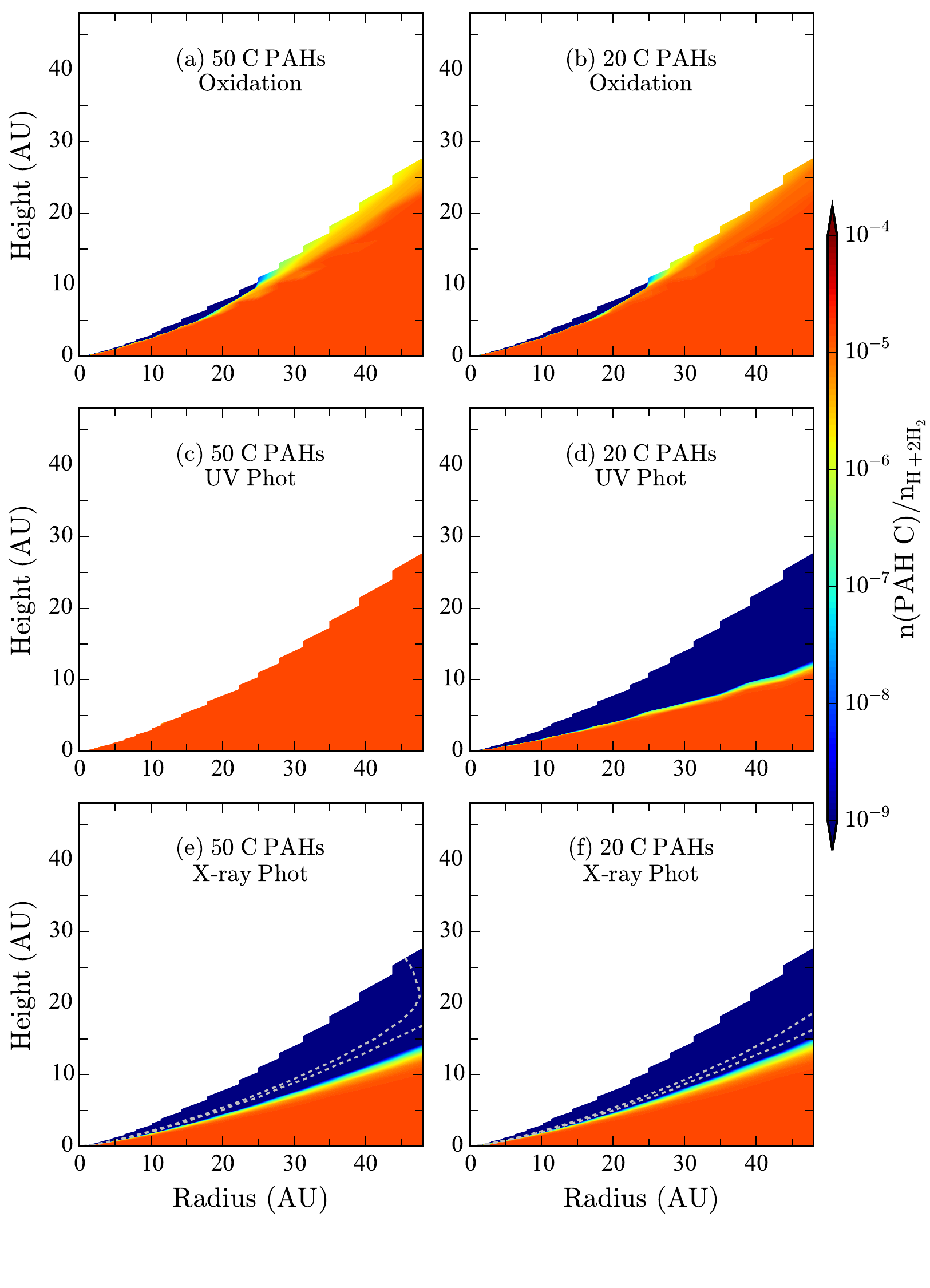}
\caption{Vertical cross-sections of the protoplanetary disk above the midplane demonstrating the refractory carbon abundance relative to total hydrogen at 10$^6$ years for PAHs containing 50 C atoms (a,c,e) and 20 C atoms (b,d,f) including oxidation (a,b), UV photodissociation using the rates of \citet[][c, d]{2007A&A...466..229V}, and X-ray photodissociation using the rates of \citet[][e, f]{2010A&A...511A...6S}. Dashed contours indicate the abundances of 10$^{-6}$ and 10$^{-9}$ for successive removal of C$_2$H by X-ray photodissociation of 50 C PAHs (e).}
\label{FinalAbundance2}
\end{figure*}

The model described in Section 2 is used to identify the locations in the disk where oxidation and photochemical destruction could cause significant depletion of refractory carbon sources. Figures~\ref{FinalAbundance}~and~\ref{FinalAbundance2} show the abundance of refractory carbon remaining after running the model for 10$^6$ years, on the order of the lifetime of a protoplanetary disk. Each panel illustrates a vertical cross-section of the disk above the midplane with contours representing the level of depletion of a particular refractory carbon source (with an abundance relative to H of 10$^{-4}$ representing zero depletion for the carbon grains and 1.5$\times$10$^{-5}$ likewise for the PAHs). Here the PAH abundance is considered to be the total abundance of all aromatic components larger than benzene in our model. As demonstrated in these panels, oxidation and photochemical destruction are only operative near the disk surface. Whereas oxidation is restricted to within a radius that depends fairly weakly on the size and nature of the refractory carbon sources, photochemical destruction can be unlimited in the radial direction out to at least 100~au.

\subsubsection{Photochemical Destruction vs. Oxidation}
In a passive disk, irradiation from the central star and the ISM produces environmental conditions---temperatures of at least a few \SI{100}{\kelvin} and high atomic oxygen abundances---in the surface layers that allow oxidation to proceed at an appreciable rate. The exponential dependence of the chemical kinetics on the gas temperature is the main limiting factor toward the disk midplane. Additionally, the dominant carriers of oxygen vary within the vertical column at each radius. Atomic O is most abundant above the cool middle layers that are protected from stellar photons, where H$_2$O ice and other molecular forms dominate, and below the surface of the disk where at radii $\lesssim$10~au ionized O$^{+}$ begins to form. Radially, at a certain distance from the star the sufficiently dense layers of the disk no longer reach the required temperatures and atomic oxygen abundances to activate oxidation and the reaction shuts off. This radial cutoff occurs at $\sim$20--65~au depending on the refractory carbon source.

Photolysis occurs over a larger region of the disk due to the lack of temperature dependence in the reaction rate. The layer to which UV photons reach extends below the hot region where oxidation can occur and to larger radii in the disk. However, the reaction is limited by the modulation of the UV field by disk material, leaving the midplane of the disk shielded. UV photolysis of HAC grains occurs down to interstellar UV levels implying the need for reformation of carbonaceous grains in the ISM (e.g., \citealp{2011A&A...530A..44J}), if they are indeed of HAC composition as predicted by current interstellar grain models \citep{2013ApJ...770...78C, 2013A&A...558A..62J}. UV photodissociation is only efficient for small PAHs, occurring within the disk lifetime for sizes $\lesssim$24 C atoms and FUV photons as shown by \citet{2007A&A...466..229V}. Significant depletion, by four orders of magnitude or greater in total refractory C abundance, occurs deeper in the disk for UV photodissociation on these small PAHs than for oxidation or X-ray photodissociation. However, the vertical cutoff for photodissociation of PAHs by X-rays is not as sharply defined as that in the UV. Minor amounts of depletion, less than 0.01\% of the total refractory C abundance, continue deeper into the disk approaching the midplane.

The timescales for UV photochemical destruction are faster than those of oxidation for all carbon grain sizes and for small PAHs (Figures~\ref{DestTime} and ~\ref{DestTime2}). In the case of the PAHs, UV photodissociation will accelerate the depletion of small PAHs following the destruction of 50 C PAHs via oxidation or X-ray photodissociation. In comparison to X-ray photodissociation, oxidation of PAHs occurs over a smaller region of the disk, but at a faster rate.

\subsubsection{Carbon Grain and PAH Sizes}
The extent of refractory carbon depletion has some dependence on the size and structure of the carbon sources. Fig.~\ref{DestTime} shows the abundance of the different carbon grains over time at radii of 1 and 10~au in the surface layers where rapid oxidation and photolysis are occurring. Larger and less porous grains are destroyed more slowly because they have a smaller [cross-section]/[occupied volume] ratio reducing the frequency of their interaction with O atoms and UV photons. For the smaller or more porous grains, faster destruction rates facilitate depletion over a larger range of vertical heights and, in the case of oxidation, radii. The cutoff in disk radius for oxidation of different grain sizes decreases from $\sim$65~au to $\sim$45~au for grain radii from 0.01 to 10 (porous)~\SI{}{\micro\metre} in panels (a), (c), and (e) of Fig.~\ref{FinalAbundance} and down to $\sim$20~au for \SI{10}{\micro\metre} compact grains. There is no such radial cutoff within 100~au for photolysis until grain sizes of \SI{10}{\micro\metre} are reached, where depletion by a factor of 100 occurs beyond 100~au but depletion by 10$^{4}$ only occurs out to 80~au. In addition, the vertical extent of depletion is slightly less for larger grains at large radii. At distances of 10s of au, depletion occurs down to $\sim$4 scale heights above the midplane for \SI{0.01}{\micro\metre} grains vs. just above 5 scale heights for porous \SI{10}{\micro\metre} grains.

Within 10$^6$ years, oxidation depletes abundances of the initial 20 or 50 C PAHs by several orders of magnitude at radii out to 100~au in the surface layers of the disk. However, this is only the first step. Breakdown of the subsequent PAH products depends on the reaction rates of each of the following steps in the network. PAH depletion occurs rapidly in the inner few au for all species. Further out in the disk where conditions become less ideal for oxidation, the results will be more dependent on the chemical pathways included in the model. For the network selected here, in 10$^6$ years, all PAH species are cleared from the surface layers as shown in Fig.~\ref{FinalAbundance2} for radii $\lesssim$30~au regardless of initial size. However, where oxidation operates, the total abundance of C in PAHs is lower when the PAHs start out in 50 C molecules as can be observed by comparing the surface layers of the disk in panels (a) and (b) of Fig.~\ref{FinalAbundance2}. Since the rate-limiting steps in the breakdown process are the destruction of the smaller, 2- or 3-ring-containing species, the refractory carbon piles up in these small PAHs. In breaking down the larger 50 C PAHs, a larger portion of the total carbon has been removed in the form of small fragments before reaching these bottleneck species. Starting with larger PAHs does slow down the rate of carbon depletion by oxidation, as shown in Fig.~\ref{DestTime2}.

UV photodissociation here only affects small PAHs. Large PAHs (greater than 24 carbon atoms, \citealt{2007A&A...466..229V}) cannot be photodissociated by FUV radiation within the disk lifetime and therefore can only be broken down by oxidation or more energetic photons. The inclusion of EUV photons would result in the break down of 50 C PAHs in addition to faster depletion of 20 C PAHs in the uppermost layers of the disk, about 1--2 scale heights above the current vertical cutoff for FUV photodissociation \citep{2010A&A...511A...6S}. X-ray photodissociation rates are slower than those for FUV photons. In the case of the complete dissociation of a PAH molecule by a single X-ray photon, destruction occurs faster (Fig.~\ref{DestTime2}) for the larger PAHs due to the dependence of the absorption cross-section on the number of carbon atoms. PAH destruction rates are decreased by an order of magnitude when considering the successive removal of C$_2$H compared to complete dissociation of the PAH molecule (Fig.~\ref{DestTime2}), which also limits the radial and vertical extent of PAH depletion (Fig.~\ref{FinalAbundance2}).

\subsubsection{Chemical Modeling Summary}
\begin{figure}
\includegraphics[scale=0.6, trim=0mm 1cm 0mm 0.6cm]{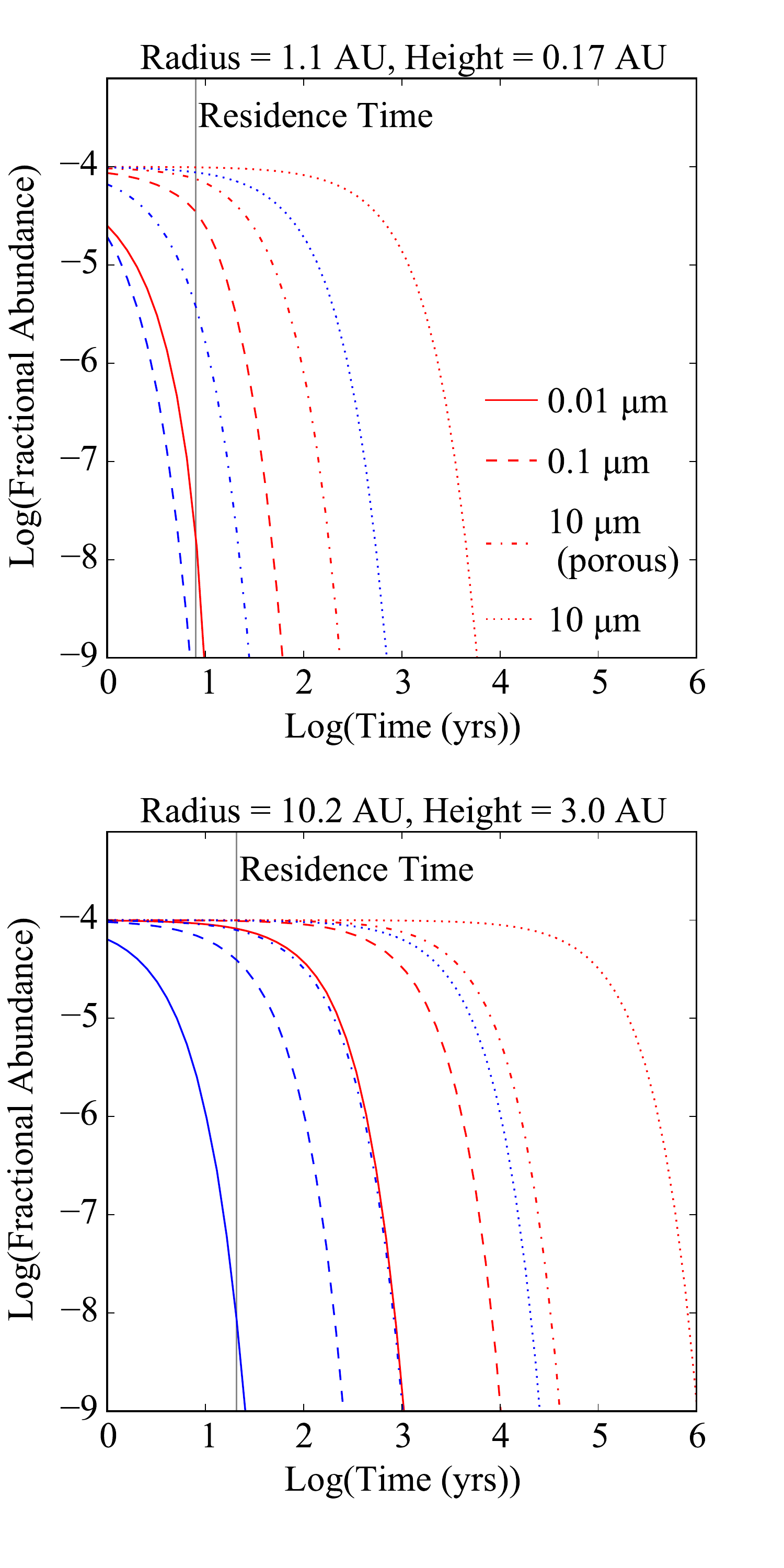}
\caption{Abundance of refractory carbon in grains over time in the middle of the destructive regions at radii of 1~au (upper plot) and 10~au (lower plot) including destruction by oxidation (red) and photolysis (blue). For comparison, the vertical line denotes the amount of time grains spend exposed to these conditions throughout the disk lifetime. At 1.1~au, the \SI{0.01}{\micro\metre} grains are depleted within the first time step due to UV photolysis.}
\label{DestTime}
\end{figure}

\begin{figure}
\includegraphics[scale=0.6, trim=0mm 1cm 0mm 0.6cm]{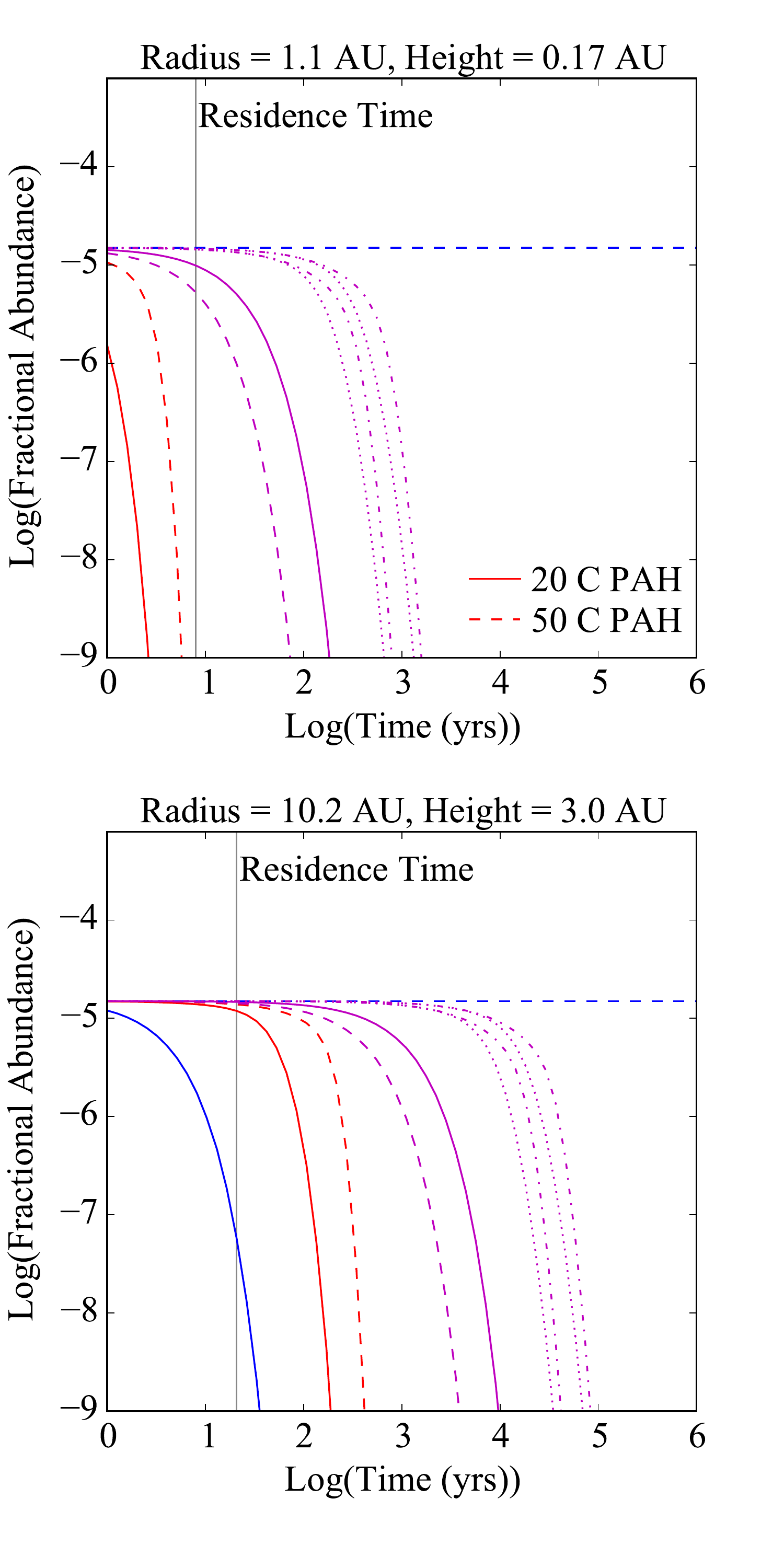}
\caption{Abundance of refractory carbon in PAHs over time in the middle of the destructive regions at radii of 1~au (upper plot) and 10~au (lower plot) including destruction by oxidation (red), UV photodissociation (blue), and X-ray photodissociation (magenta). For comparison, the vertical line denotes the amount of time grains spend exposed to these conditions throughout the disk lifetime. At 1.1~au, the 20 C PAHs are depleted within the first time step due to UV photodissociation. Rates of X-ray photodissociation are shown for entire PAH destruction by individual X-rays (solid lines for 20 C PAHs, dashed lines for 50 C PAHs) and successive removal of C$_2$H by X-rays with a yield of 1.0 and 0.5 (dotted lines for 20 C PAHs, dash-dotted lines for 50 C PAHs).}
\label{DestTime2}
\end{figure}

Figures~\ref{DestTime} and~\ref{DestTime2} demonstrate that under favorable conditions present in the surface layers of the disk, depletion of refractory carbon sources occurs very rapidly compared to the disk lifetime. In terms of the chemistry, the amount of refractory carbon destroyed in these regions can exceed that required to explain the disparity among carbonaceous chondrites and the Earth relative to the interstellar dust (1--2 and $\sim$4 orders of magnitude, respectively). As shown in Fig.~\ref{FinalAbundance}~and~\ref{FinalAbundance2}, these carbon-deficient regions in the surface layers reach out beyond the terrestrial-planet- and asteroid-forming regimes of the disk.  However, material in the midplane remains largely unaffected by oxidation and photochemical destruction in the static chemical model.

\subsection{Radial Extent of Refractory Carbon Depletion}

The conditions required for oxidation and photochemical destruction become less attainable (higher up in the disk) at larger distances from the central star suggesting that there may be a radial cutoff beyond which the mechanism is ineffective and refractory carbon will remain in the condensed phase at abundances similar to the interstellar value. In contrast to snow lines, this boundary, akin to the ``soot line" described by \cite{2010AdSpR..46...44K} for the abundance of PAHs, would mark the location of an irreversible transition. Once carbon in the inner disk enters the gas phase it will likely remain volatile given the limited mechanisms available to return it to a more-refractory state. The high abundance of refractory carbon in the ISM will cause a potentially drastic change in the carbon chemistry from one side of the transition region, where nearly all the cosmically available carbon would be in volatile form and---in the warm inner disk---largely removed from the planetesimal formation process, to the other, where about half of this carbon would exist in a more-refractory phase and be available to be incorporated into forming planetesimals. Therefore, estimating the radial extent of carbon depletion mechanisms such as oxidation and photochemical destruction may allow us to relate the chemistry of the disk to the carbon content of solar-system bodies. 

\begin{figure}[t]
\centering
\includegraphics[scale=0.9, trim=1cm 0cm 0mm 0cm]{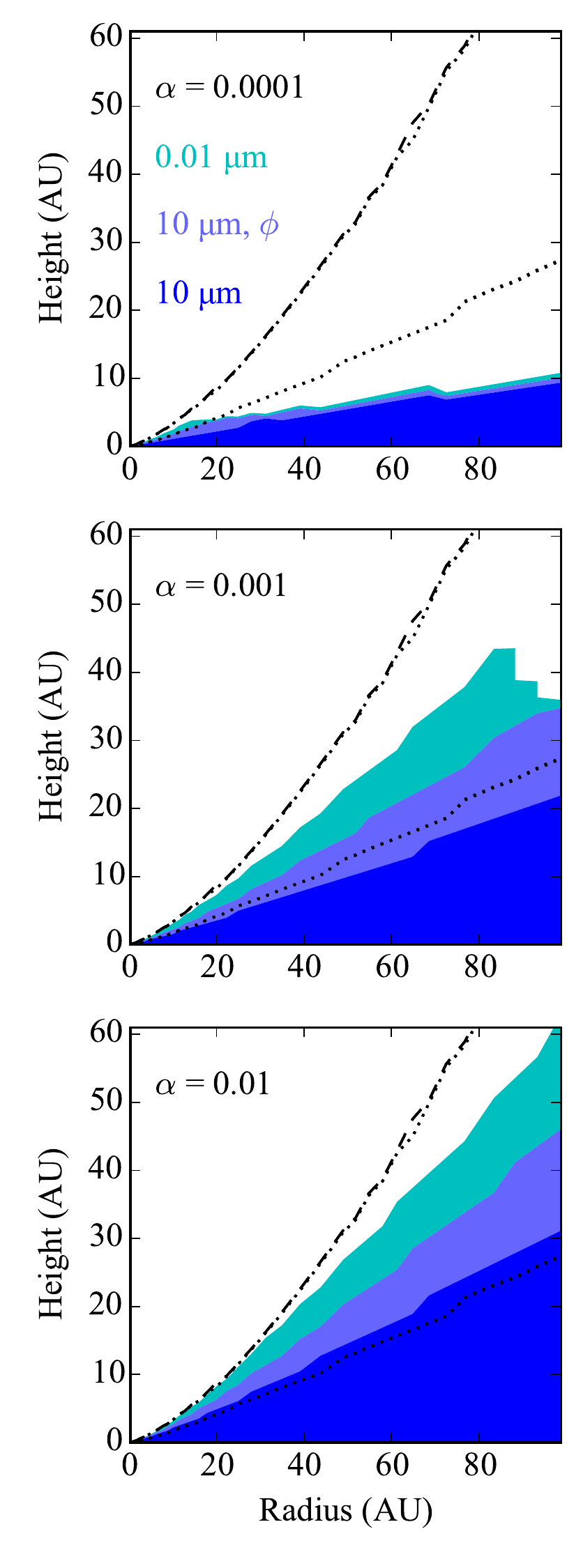}
\caption{Maximum height to which carbon grains of sizes 0.01 and \SI{10}{\micro\metre}, porous ($\phi$) and compact, can be lofted for different levels of disk turbulence parameterized by $\alpha$ = $10^{-4}$--$10^{-2}$. For reference, the height of the UV destructive layer for \SI{0.01}{\micro\metre} carbon grains is indicated by the dotted line and the disk surface by the dash-dotted line.}
\label{MaxHeight}
\end{figure}

As shown in Section 3, active oxidation and photochemical destruction of refractory carbon is limited to the photochemically active surface layers. However, the majority of the disk material is concentrated in the midplane and it is in this region that planet formation occurs. As suggested by \cite{2010ApJ...710L..21L}, mixing within the disk could introduce dust from these deeper layers into the destructive regions causing depletion throughout the vertical column. Turbulent motion is thought to counteract dust grain settling, continually stirring up disk material and maintaining the presence of grains in layers high above the midplane \citep{, 1993prpl.conf.1031W, 2004A&A...421.1075D, 2010ApJ...723..514C}. Fully assessing the extent to which destruction in the surface layers could affect material in the midplane requires a treatment of dust evolution and transport including processes such as growth, settling, fragmentation, and radial drift. The extent will further depend upon the underlying mechanism driving disk accretion and causing the turbulent motions throughout the disk. The dominant radial flow in the disk midplane is inward, but there is also evidence for outward transport in the solar nebula (e.g. crystalline silicates found in the comet Wild~2 by the NASA Stardust mission, \citealp{2006Sci...314.1711B}) that could carry carbon-depleted materials from close to the Sun to larger radii. Given these complexities, here we start by only considering the limit where vertical motion is more rapid than this radial transport and where the turbulence in the disk is characterized by a constant value of the Shakura \& Sunyaev $\alpha$ parameter ranging from $10^{-4}$ to $10^{-2}$ \citep{1998ApJ...495..385H, 2012A&A...539A...9M}.

The vertical height to which grains can be suspended due to turbulent motion is estimated by comparing the timescale of grain stirring to that of grain settling throughout the disk as described by \cite{2004A&A...421.1075D} and used by \cite{2010ApJ...710L..21L}. The stirring timescale ($t_{stir} =  \mathrm{Sc}(z) z^{2}/[\alpha c_{s} H]$) describes the amount of time required for grains to diffuse to a height $z$ above the midplane, effectively describing the time required for components within a vertical column between the midplane and $z$ to become well-mixed. The Schmidt number, $\mathrm{Sc}(z)$, is assumed to equal one such that the small particles considered here are perfectly coupled to the turbulence. This number can greatly exceed one in the low-density, upper layers of the disk making the stirring timescale used here a lower limit. In addition to the height above the midplane, $t_{stir}$ depends on the level of turbulence parameterized by $\alpha$ (a more turbulent disk has shorter stirring timescales) and the temperature structure of the disk, which affects the sound speed $c_{s}(z)$ and disk scale height $H$ = $c_{s}(0) / \Omega_{K}$, where $\Omega_{K}$ is the Keplerian rotation rate at radius $R$. The settling timescale ($t_{sett} = [4 \sigma c_{s}(z) \rho_{gas}(z)]/[3 m \Omega_{K}^2]$) is shorter for larger and more compact grains due to its dependence on the cross-section/mass ratio [$\sigma / m = 3 / (4 \rho_{cgr} (100\%-\phi) r)$] of the grains but is also affected by the disk temperature and density [$\rho_{gas}(z)$] structure. Grains can be lofted up to a maximum height, where $t_{stir}$ = 100 $\times$ $t_{sett}$, above which the downward gravitational force sufficiently overcomes the turbulent motion causing grains to be completely settled out. 

Figure~\ref{MaxHeight} shows the maximum height reached by grains of sizes 0.01--\SI{10}{\micro\metre} within the disk lifetime given the conditions in our physical disk model. Due to their longer settling times, smaller and more porous grains can reach heights closer to the disk surface. Excluding the larger (1--\SI{10}{\micro\metre}) compact grains, all other grains can be lofted into the photo-destructive layer shown in Fig.~\ref{FinalAbundance} out to radii of nearly 100 or more au for sufficiently turbulent disks ($\alpha$ = $10^{-3}$--$10^{-2}$).  This distance is only $\sim$10--100+~au for compact \SI{1}{\micro\metre} grains and 0--2~au for compact \SI{10}{\micro\metre} grains for the same range in $\alpha$. At low $\alpha$, $t_{stir}$ approaches the disk lifetime of several million years and therefore restricts the upward mobility of grains of all sizes. For $\alpha$ = $10^{-4}$, only grains smaller or more porous than compact \SI{1}{\micro\metre} grains can be lofted into the photo-destructive layers and within a limited distance from the star of $\sim$15~au.

PAHs in the gas phase would remain coupled to the gas, similar to the small carbon grains, and are not subject to settling. However, given that hydrocarbons have a sublimation temperature of $\sim$\SI{400}{\kelvin} \citepalias{2015PNAS..112.8965B}, in dense regions of the disk where the chance of interaction with grains is high, PAHs will likely freeze out and remain on grain surfaces. We would expect these PAHs to attach mainly to the smallest grains because they represent the largest surface area. Therefore, in the limit where all of the PAHs are frozen out, the results would be similar to those of the \SI{0.01}{\micro\metre} grains. PAHs that evade interaction with grains would be destroyed at the rates presented in Section 3.1 at the disk surface where conditions are appropriate for oxidation, but would remain largely unaffected in deeper layers of the disk. 

Even if grains are able to reach the heights where destruction occurs within the disk, the amount of time grains spend at those heights is equally important to ensure their destruction. Where the stirring timescale is short compared to the disk lifetime, grains will cycle through the vertical column multiple times. In lower $\alpha$ disks with longer stirring timescales, grains tend to experience longer excursions through a smaller number of different environments. Alternatively, in higher $\alpha$ disks with shorter stirring timescales, grains rapidly change environments spending brief intervals of time in each one \citep{2010ApJ...723..514C}. The average time spent in a particular region over multiple stirring cycles ends up being only weakly dependent on $\alpha$, however, and related primarily to the density structure of the disk. Here the fraction of the disk lifetime that an average grain spends above a height $z$ is taken to be approximately the ratio of the integrated gas density above height $z$ to the total gas density of the vertical column above the midplane.

\begin{figure}[t]
\centering
\includegraphics[scale=0.58]{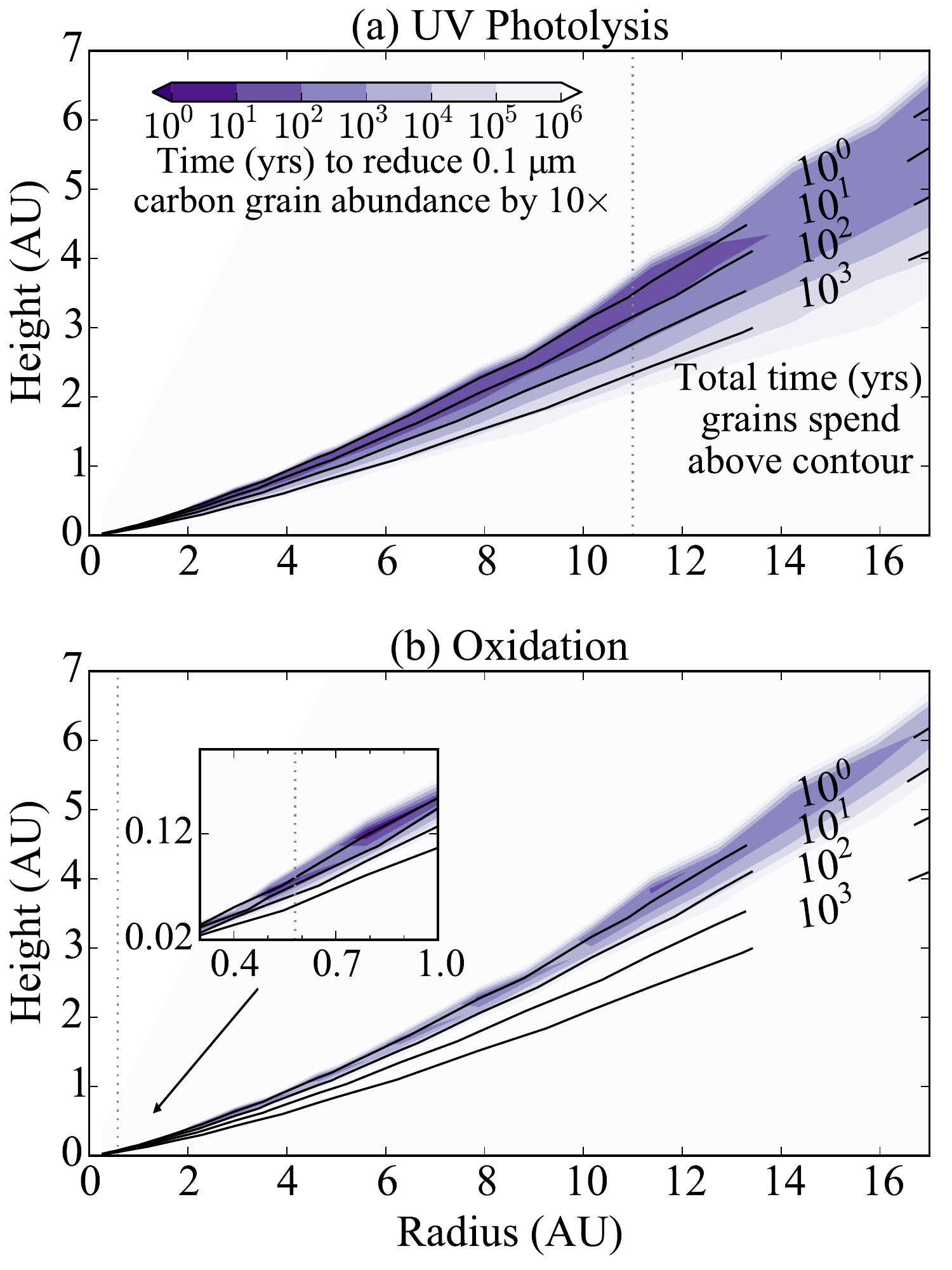}
\caption{Comparison of the time grains are exposed to the conditions above heights outlined in contours to the time required to deplete the abundance of \SI{0.1}{\micro\metre} carbon grains by an order of magnitude via UV photolysis (a) and oxidation (b) represented by the filled contours. The dotted line represents the radius beyond which the depletion time is less than the exposure time and the \SI{0.1}{\micro\metre} carbon grains will mostly survive. As discussed in the text, the exposure time does not strongly depend on the choice of $\alpha$ regarding the disk turbulence.}
\label{Exposure}
\end{figure}

 Figure~\ref{Exposure} compares the time spent at heights above the midplane for a disk lifetime on the order of $10^6$ years to the time required to deplete the \SI{0.1}{\micro\metre} grains by one order of magnitude. Although carbon grain destruction occurs rapidly, grains would only be exposed to destructive conditions several scale heights above the midplane for $\sim$1--10 years for oxidation or $\sim$100--1000 years for photolysis. Where the depletion timescale, represented by the filled contours in Fig.~\ref{Exposure}, is longer than the exposure timescale and therefore a filled contour crosses over solid contour lines corresponding to comparable or shorter amounts of time, depletion of refractory carbon in \SI{0.1}{\micro\metre} grains can occur within the disk lifetime. For UV photolysis in panel (a), the crossover occurs around 10~au, where the depletion and exposure times are on the order of 10 years. In comparison, this crossover occurs at $\sim$0.5~au for oxidation. These numbers represent upper estimates because they assume that the particles spend all of their time behaving aerodynamically as small monomers.  This may be true in very turbulent cases, where collisions become destructive and prohibit growth or if grains are part of very porous aggregates.  Growth to large or more compact aggregates may limit the time spent at these higher altitudes \citep{2016ApJ...822..111K}.

\begin{figure*}
\centering
\includegraphics[scale=0.43, trim=1cm 0cm 0mm 0cm]{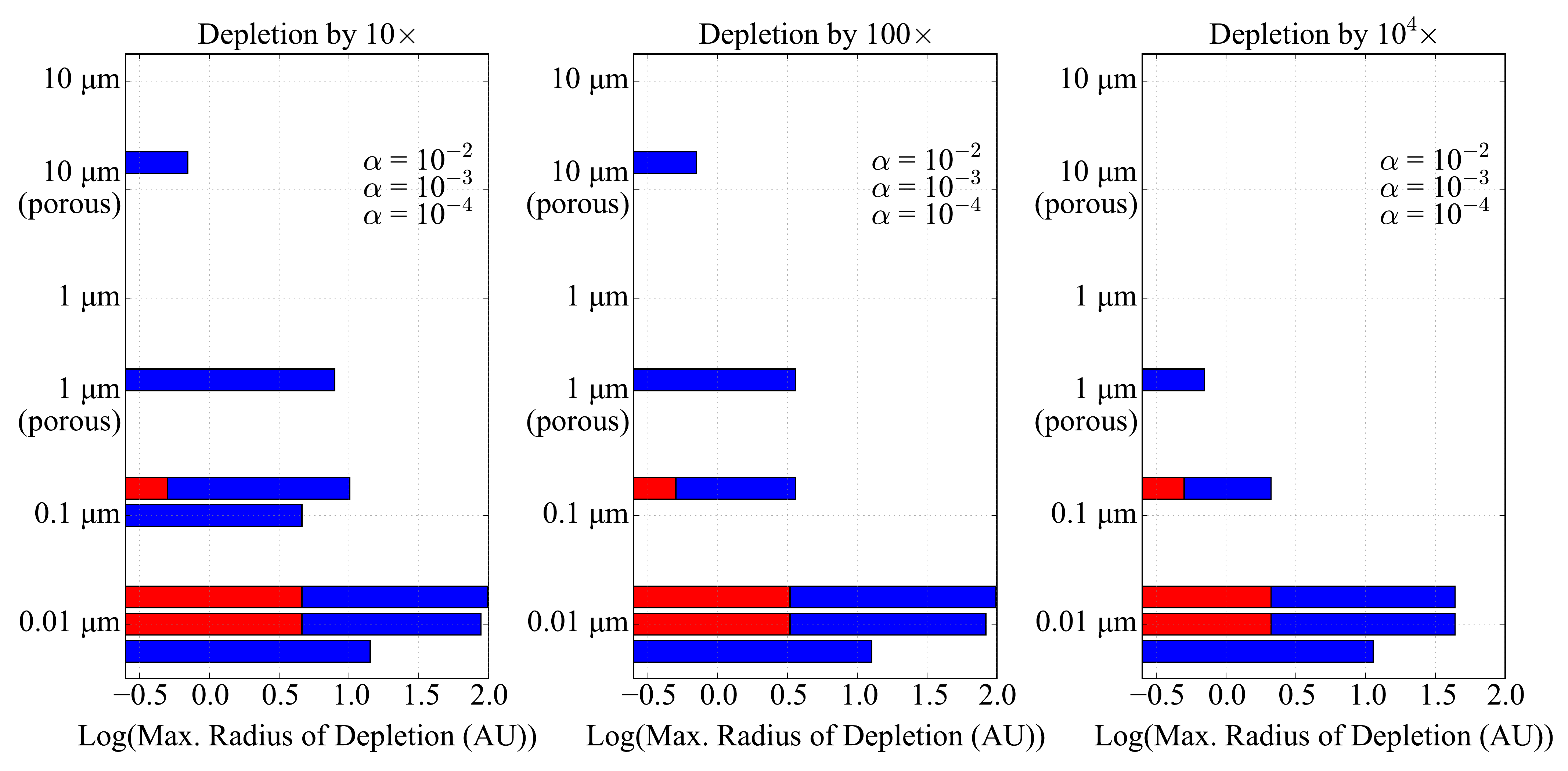}
\caption{Maximum radii where depletion in the midplane can occur---where grains can be turbulently lofted to a height multiple times within the disk lifetime and the time to deplete the carbon grain population is shorter than the total fraction of the disk lifetime that grains spend at that height---for different grain sizes, levels of turbulence, and levels of carbon abundance depletion including oxidation (red) and photolysis (blue). }
\vspace{0.1cm}
\label{New}
\end{figure*}

In the exposure time, refractory carbon abundances can be depleted to the levels observed in solar-system bodies out to maximum radii shown in Figure~\ref{New}. For \SI{0.01}{\micro\metre} grains, the maximum is about 3--5~au for depletion levels similar to the carbonaceous chondrites via oxidation. However, UV photolysis will clear out these grains throughout the planet-forming region. Larger, from compact 0.1 to porous \SI{1}{\micro\metre}, grains are depleted out to less than 1~au via oxidation but 3--10~au via photolysis. Once carbon has been incorporated into grains of \SI{10}{\micro\metre} or larger (or compact \SI{1}{\micro\metre} grains), it will likely survive these processes and remain in refractory form with the exception of porous grains in very turbulent disks within 1~au.  For a constant-$\alpha$ disk, depletion requires sufficient turbulence to allow for lofting of the grains into the surface layers within the disk lifetime. Values of $\alpha \sim$10$^{-2}$ are needed to deplete carbon from \SI{1}{\micro\metre} grains out to even a few au. For lower turbulence disks, depletion is limited to submicron grain sizes out to a few au ($\alpha$=10$^{-3}$) or \SI{0.01}{\micro\metre} grains out to $\sim$10~au ($\alpha$=10$^{-4}$).

The sizes of the depleted regions in the surface layers of the disk from Fig.~\ref{FinalAbundance} represent the upper limit for the radial extent of carbon depletion assuming perfectly efficient transfer of material from the midplane to the surface and sufficient exposure times to destruction mechanisms. The actual extent of depletion in the midplane will almost certainly be less and highly dependent on the transport processes, grain evolution, and the temperature and density structure within the disk. In the simplest case of a constant-$\alpha$ model with our physical disk structure, depletion of submicron to micron-sized carbon grains can occur within $\lesssim$3--10~au in turbulent disks and average interstellar-sized carbon grains can be cleared out of the planet-forming region.

\section{Discussion}
The analysis of the potential refractory carbon depletion due to oxidation and photochemical destruction depends on several unconstrained factors. The extent to which grains can be destroyed depends on their growth rate. Smaller grains are destroyed faster and can be lofted higher allowing them to reach the surface layers more easily. Larger grains are more likely to survive destruction. Therefore, the effectiveness of this destruction mechanism depends on the timescale and efficiency of grain growth. When growth is accompanied by collisional erosion and fragmentation, maintaining a constant supply of small grains in the disk, photochemical destruction and combustion remain effective. However, the formation of sufficiently large bodies that sweep up small dust grains prevents the small grains from traveling efficiently from the midplane to the surface \citep{2016ApJ...822..111K} and therefore hinders the destruction of refractory carbon.

The locations where the tested destruction mechanisms occur at an appreciable rate depend on the radiation field and temperature structure (for oxidation) within the disk. In order to have sufficient UV photons for photochemical destruction of refractory carbon in disks around low-mass stars where the UV field is dominated by accretion luminosity, the star must be accreting. The current analysis is based on an approximately solar-mass star. Increasing the stellar mass and therefore its luminosity would result in a warmer disk and potentially extend the distances out to which destruction can occur. However, changing the stellar mass would alter the input stellar irradiation field as well, which is a topic to be explored in future work. The assumed disk properties are also important. Younger disks may be more massive, preventing radiation from penetrating as deeply into the disk affecting both the temperature structure and atomic O abundances, causing the photoactive and oxidative regions to be higher above the midplane away from the majority of the disk material. An active disk undergoing accretion and potentially subject to large bursts of episodic accretion similar to those of FU Orionis objects will have a vastly different temperature profile than the passive disk modeled here. Warming dense regions of the disk may increase the effectiveness of oxidation throughout the disk. 

Furthermore, this analysis depends on the composition, optical properties, and distribution of materials within the disk. Dust grains may not have pure carbon surfaces and in cooler regions of the disk may be coated in volatile ice. This would reduce the efficiency of destruction mechanisms tested here. The change in opacity due to icy mantles could also alter the modeled temperature distribution in the disk. However, evidence suggests that ice coatings may not be a concern for this analysis. Observed water emission due to UV desorption of water ice from grains in the outer disk is relatively weak. This may be the result of differential settling where larger, ice-coated grains typically reside below the UV-irradiated layers and the small grains above have bare surfaces \citep{2011Sci...334..338H}. Additional settling of dust grains, approximated by increasing the fraction of the dust mass in larger grains and restricting the large grain population to lower scale heights, has little effect on the location of the high temperature ($>$\SI{100}{\kelvin}) gas in our model.

Mass transport within the disk is required in order for this mechanism to reproduce the observed carbon deficit in solar-system bodies relative to interstellar dust. In the constant-$\alpha$ model, sufficiently turbulent disks are able to loft grains above the midplane. However, increasing the amount of time grains spend exposed to destructive conditions requires some asymmetry in their vertical motion that would cause them to spend additional time in the upper layers of the disk. Consideration of different angular momentum transport processes may be important for this analysis. For example, a disk model including wind-driven accretion, where \cite{2013ApJ...769...76B} found that at 1~au the accretion flow occurs in a thin layer offset from the midplane by multiple scale heights, may alter the patterns of dust migration relative to the destructive regions. To further explore grain motion, a numerical (i.e. non-ideal) turbulent disk model could be employed, coupling dust evolution to the physical and chemical state of the disk and allowing timescales and lofting heights to be determined by averaging over the trajectories and lifetimes of individual grains in the simulation \citep{2010ApJ...723..514C, 2011ApJ...740....9C}. Ultimately, such models could be combined with dust coagulation and planetesimal formation scenarios to provide a quantitative assessment of the refractory carbon distribution in the disk prior to and through the formation of planetesimals. 

\section{Conclusion}
We ran a chemistry model for an irradiated, passive disk, including destruction mechanisms for solid carbon grains and PAHs -- two potential sources of refractory carbon that could have been inherited from the ISM and present in the protoplanetary disk. Oxidation and photochemical destruction rapidly deplete refractory carbon but are limited to the photochemically active surface layers of the disk. Oxidation and photolysis of large grains are further limited within a particular radial distance depending on the size and structure of the refractory carbon source. The maximum radial distance to which refractory carbon can be oxidized at the surface is $\sim$20--65~au after 10$^6$ years for the 0.01--\SI{10}{\micro\metre} grains and $\sim$30~au for PAHs whereas photochemical destruction can extend out to 100+~au.

Motion of grains within the disk is required to deplete refractory carbon at the midplane to the levels observed in solar-system bodies. This motion is difficult to model analytically. Approximations for the timescales of the average motion of dust grains are used to constrain the extent to which refractory carbon can become depleted at the midplane. For our model of a passive, constant-$\alpha$ disk with high turbulence, carbon grains smaller or more porous than compact \SI{10}{\micro\metre} grains (and PAHs frozen to their surfaces) can be lofted into the destructive regions within 10--100+~au from the central star but their depletion is limited by the amount of time they are exposed to the destructive conditions. UV photolysis has proved to be an important mechanism in depleting refractory carbon. The fast reaction rates allow for destruction of grains in the surface layers of the disk even if they spend very little time there and the lack of temperature dependence extends the destructive region to any surface layers with sufficient UV radiation.

Early on in planet formation, cm-sized and larger ``pebbles" are built up through the interaction of smaller grains starting with interstellar sizes and compositions. These initial small grains will be subjected to physical and chemistry processes in the protoplanetary disk. While these grains remain small, photolysis (and to a lesser extent oxidation) can selectively erode the refractory carbon component of the population releasing it into the volatile phase in the inner portions of the disk. Photolysis can destroy the  \SI{0.01}{\micro\metre} carbon grains throughout the planet-forming region of the disk assuming sufficient turbulence. However, the 0.1--\SI{1}{\micro\metre} grains may be a more significant source of material for planetesimals. If the primordial grain size distribution is similar to that of the ISM, small grains will represent most of the surface area but the bulk of the mass will be in the 0.1--\SI{1}{\micro\metre} grains. Once grains reach \SI{10}{\micro\metre} in size they will be largely unaffected by oxidation and photochemical destruction unless they are broken down and rebuilt from smaller grains. Therefore, in order to be effective, destruction of refractory carbon grains will need to occur early in the lifetime of the disk prior to significant grain growth and the building of planetesimals.

Based on our model, submicron to micron-sized carbon grain abundances can be depleted down to the levels of the carbonaceous chondrites and planetesimals sampled in the atmospheres of polluted white dwarfs out to a few to 10~au in sufficiently turbulent disks. Interstellar-sized grains can be cleared out of the planet-forming region of the disk in such turbulent disks and up to $\sim$10~au even in disks with lower turbulence. However, this analysis depends on several unconstrained parameters in the disk including the temperature and density structure, the amount of turbulence present, and the nature of the carbon sources. Further exploration of refractory carbon depletion in protoplanetary disks may therefore require consideration of alternative disk structures, dust transport, and/or accretion mechanisms. Ultimately, estimating the efficacy and radial cutoff of refractory carbon destruction mechanisms within the protoplanetary disk, including oxidation and photochemical destruction, may provide an explanation for the carbon content of planetary bodies in our solar system and how it relates to their place of origin.

\acknowledgments
The authors thank the anonymous reviewer whose comments and suggestions improved this work. This material is based upon work supported by the National Science Foundation, via the Graduate Research Fellowship Program under Grant No. DGE-1144469 and the Astronomy and Astrophysics Research Grants Program under Grant No. AST-1514918.

\bibliography{AndersonDana_bibliography}

\newpage
\appendix
\setcounter{figure}{0}
\renewcommand{\thefigure}{A\arabic{figure}}
Regions of the disk where active depletion of refractory carbon occurs are determined by the physical conditions present. Figure~\ref{App1} aids in the direct comparison of the abundance of refractory carbon after 10$^6$ years for models shown in  Figures~\ref{FinalAbundance}~and~\ref{FinalAbundance2} to the gas temperature and radiation fields from Figure~\ref{Disk} for select radii.
\vspace{-2cm}
\begin{figure}[!b]
\centering
\includegraphics[scale=0.8, trim=0cm 1.6cm 0cm 2cm]{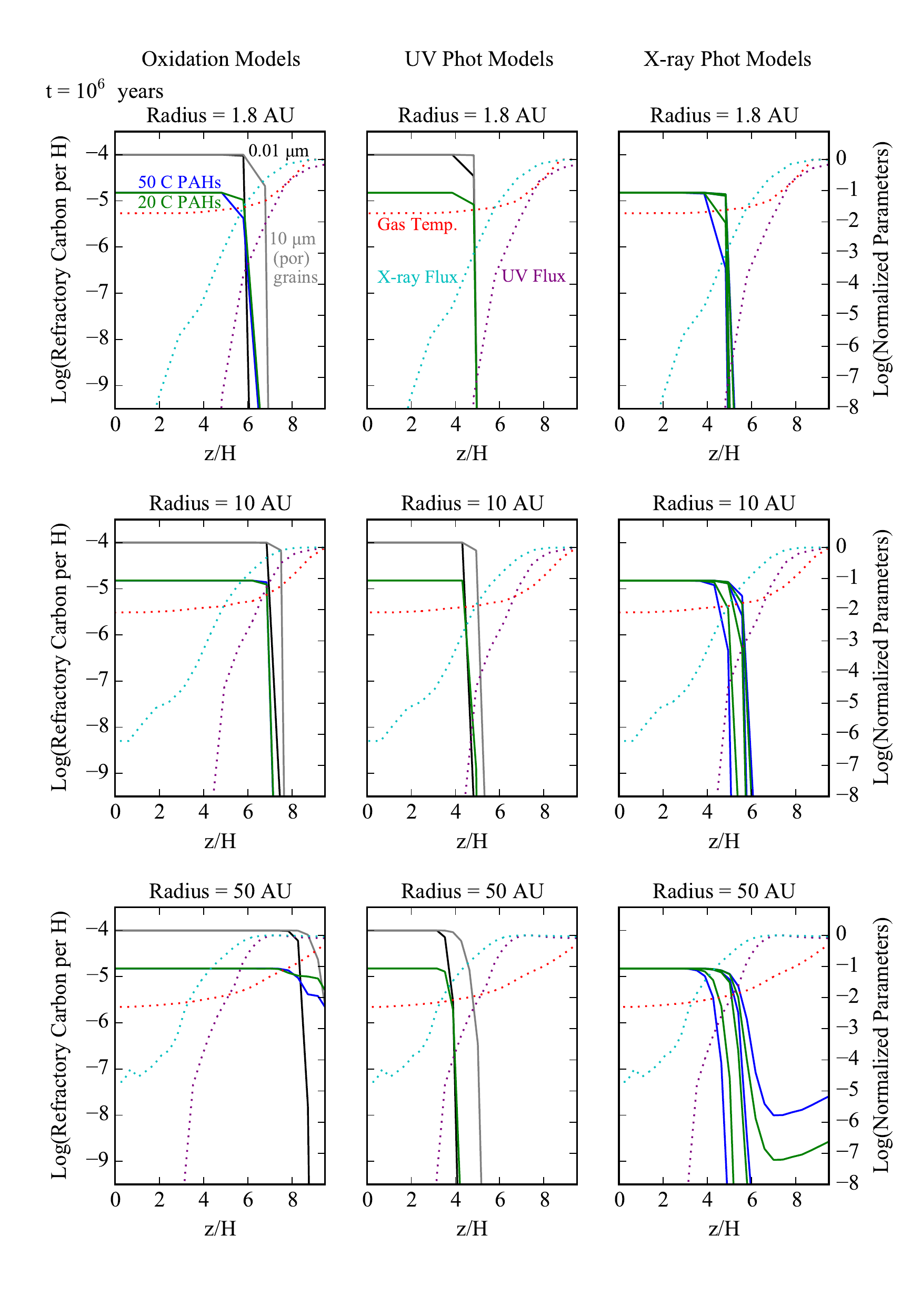}
\caption{ Vertical profiles relative to the disk scale height (H) of refractory carbon abundance (solid lines) after 10$^6$ years on the left axis and gas temperature (red dotted line), integrated UV flux (purple dotted line), and integrated X-ray flux (cyan dotted line) normalized to the maximum value for each radius on the right axis. Models of oxidation, UV and X-ray photodissociation of 0.01 and porous \SI{10}{\micro\metre} carbon grains (black and gray solid lines, respectively) and 50 and 20 C PAHs (blue and green solid lines, respectively) are shown. See Section 2.3 for further description of these models. Maximum values used for normalization are \SI{4200}{\kelvin}, 2.9$\times$10$^6$~G$_0$, and 4.8$\times$10$^9$~photons cm$^{-2}$ s$^{-1}$ at 1.8~au; \SI{4200}{\kelvin}, 6.6$\times$10$^4$~G$_0$, and 1.4$\times$10$^9$~photons cm$^{-2}$ s$^{-1}$ at 10~au; and \SI{4200}{\kelvin}, 2.5$\times$10$^3$~G$_0$, and 4.9$\times$10$^7$~photons cm$^{-2}$ s$^{-1}$ at 50~au.}
\label{App1}
\end{figure}

\end{document}